# Using Noise to Augment Synchronization among Oscillators

Jaykumar Vaidya, Mohammad Khairul Bashar, Nikhil Shukla*

Department of Electrical and Computer Engineering, University of Virginia, Charlottesville, VA- 22904 USA

*e-mail: ns6pf@virginia.edu

**Abstract**

**Noise is expected to play an important role in the dynamics of analog systems such as coupled oscillators which have recently been explored as a hardware platform for application in computing. In this work, we experimentally investigate the effect of noise on the synchronization of relaxation oscillators and their computational properties. Specifically, in contrast to its typically expected adverse effect, we first demonstrate that a common white noise input induces frequency locking among uncoupled oscillators. Experiments show that the minimum noise voltage required to induce frequency locking increases linearly with the amplitude of the oscillator output whereas it decreases with increasing number of oscillators. Further, our work reveals that in a coupled system of oscillators – relevant to solving computational problems such as graph coloring, the injection of white noise helps reduce the minimum required capacitive coupling strength. With the injection of noise, the coupled system demonstrates frequency locking along with the desired phase-based computational properties at 5× lower coupling strength than that required when no external noise is introduced. Consequently, this can reduce the footprint of the coupling element and the corresponding area-intensive coupling architecture. Our work shows that noise can be utilized as an effective knob to optimize the implementation of coupled oscillator-based computing platforms.**

**Introduction**

Coupled oscillators have experienced renewed interest in computation owing to their rich spatial-temporal properties[1,2]. Besides their potential application in realizing associative memory[3,4] and oscillator neural networks (ONNs)[5–8] for tasks such as image processing[9,10], these systems have recently been explored for solving hard combinatorial optimization problems which are still considered intractable to solve using conventional digital computers. Examples of such problems include graph coloring[11] (representative problem considered here), computing the maximum independent set[12] and maximum cut of a graph[13-16] among others. While coupled oscillators can provide an alternate, and potentially more efficient, pathway to solving such problems, one of the important factors in the design and implementation of such analog systems is noise. Normally, the injection of external noise should have adverse effects on the performance of electronic circuits with analog circuits such as oscillators being particularly susceptible. In fact, this was an important consideration in the adoption of digital circuits over analog ones in the 1950s[17].

However, in contrast to its typically undesirable effects, noise can play a constructive role in promoting the highly non-linear process of synchronization among oscillators. Prior work has studied the effects of different types of noise on the synchronization of oscillators, both, theoretically (example, [18–27]) and experimentally (example, [28–30]); identical and non-identical oscillators have been shown to exhibit synchronization in response to both white and colored noise. Noise induced synchronization has been explored in various neural networks [31-33]. In fact, the effect of noise on synchronization has even been explored in biological systems such as spike generation in neurons of neocortical slices of rats[34], firing patterns of two uncoupled neurons in paddlefish[35], as well as in other physical systems such as lasers[36,37]. However, the effect of noise injection on the synchronization of oscillators in the context of their computational properties, particularly for solving combinatorial optimization problems (here, graph coloring), remains largely unexplored. Therefore, in this work, we investigate using experiments and simulations, the role of noise in the coupling dynamics of oscillators and on their resulting computational properties. Specifically, we demonstrate that the injection of noise lowers the minimum coupling strength required to induce frequency locking and the subsequent phase properties required for computation.

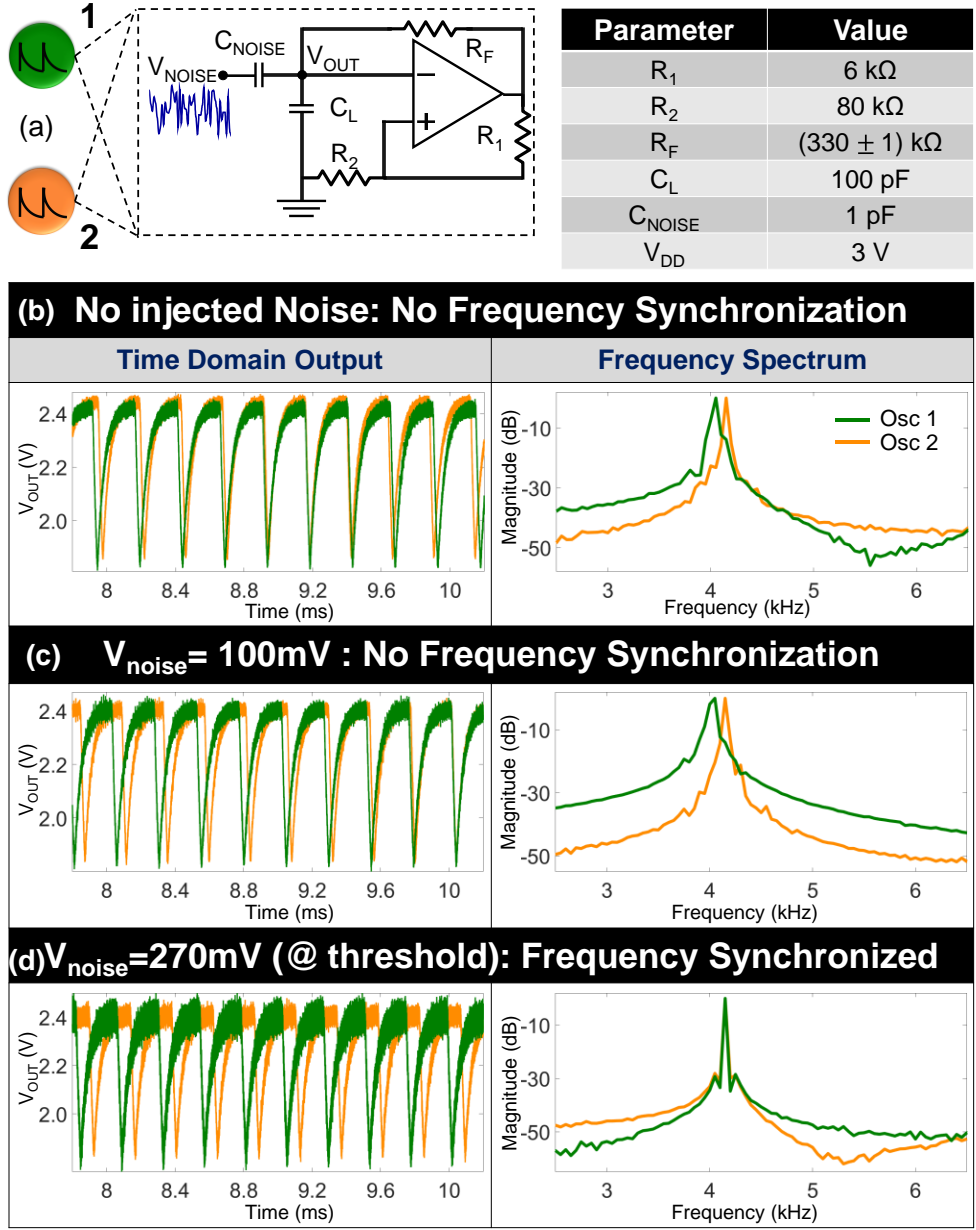

Figure 1. **(a)** Schematic of the Schmitt-trigger oscillator along with the values of the components used in the experimental realization. Time domain output and frequency spectrum of the two oscillators for various white noise inputs: **(b)** No noise; **(c)** $V_{NOISE}$=100 mV$_{RMS}$; **(d)** $V_{NOISE}$ = 270 mV$_{RMS}$. No frequency locking is observed until $V_{NOISE} \geq$ 270 mV$_{RMS}$.

**Results**

**Synchronization of two uncoupled oscillators.** We first investigate the effect of noise on the synchronization of two uncoupled oscillators with a common (white) noise input. Fig. 1a shows the schematic of a Schmitt-trigger based relaxation oscillator along with the component values used in the discrete (breadboard-based) experimental realization; the Schmitt-trigger is designed using an OPAMP (LM741) and the oscillations are stabilized using a negative RC feedback. The frequency of oscillations can be tuned using the resistor ($R_F$) and the capacitor $C_L$ in the negative feedback loop; we intentionally introduce a small change in $R_F$ of the two oscillators to ensure they have slightly different frequencies and are not synchronized trivially. White noise generated using a function generator (Keysight 81160A) is injected at the output of the oscillator through a capacitor ($C_{NOISE}$= 1pF); the value $C_{NOISE}$ is chosen such that the oscillators, in the absence of noise, do not exhibit frequency locking using only $C_{NOISE}$. The oscillator outputs are measured using a digital oscilloscope (Keithley DSO104A).

Figure 1b-d shows the time domain waveforms and the corresponding frequency spectrum of the oscillators when subjected to different levels of external white noise. In the absence of externally injected noise (Fig. 1b), no frequency locking is observed among the oscillators. Furthermore, the oscillators fail to frequency lock (Fig. 1c)

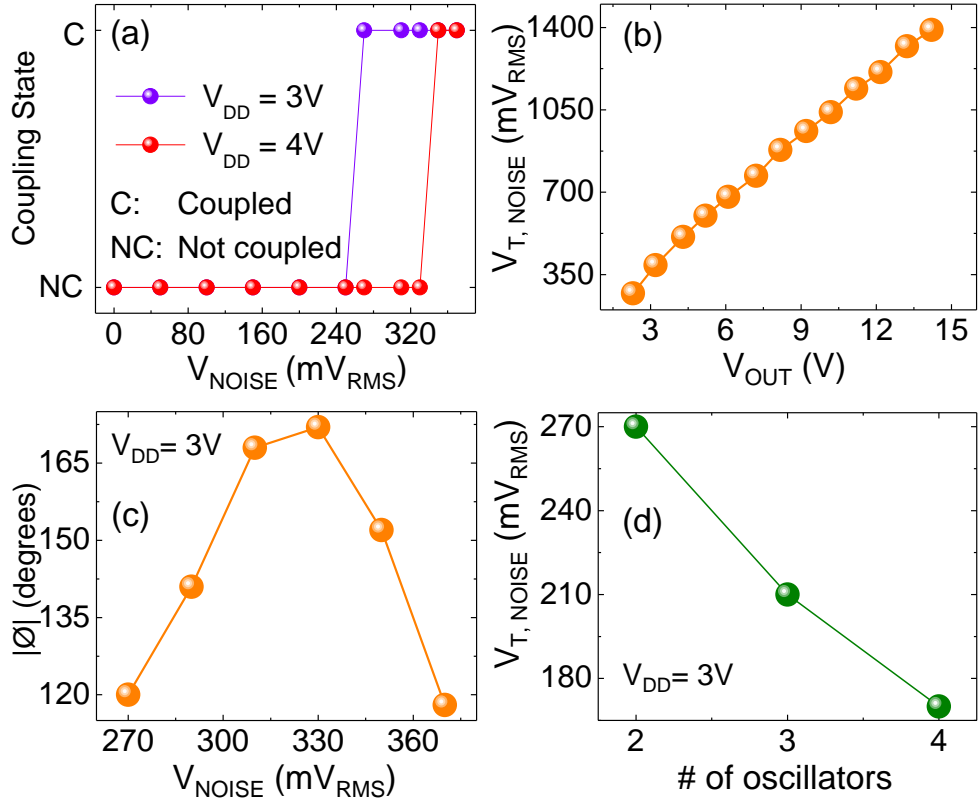

**Figure 2. (a)** Effect of noise voltage on the synchronization of the two oscillators for two different biases ($V_{DD}$=3V, 4V). **(b)** Variation of noise threshold (for frequency locking) with amplitude of the output. **(c)** Evolution of the relative phase difference (Ø) between the synchronized oscillators as function of the noise voltage. **(d)** Minimum noise threshold ($V_{T,NOISE}$) required for synchronization as a function of the number of oscillators; the threshold decreases with increasing number of oscillators.

until the noise voltage reaches the threshold value of 270 mV$_{RMS}$ (Fig. 1d) even though the frequency mismatch among the oscillators decreases as the intensity of the injected noise is increased. However, when the noise voltage equals or exceeds this threshold input, the oscillators exhibit (noise induced) frequency locking as evident by the fact that the resonant peaks coalesce to the same frequency in the spectrum. Moreover, the frequency locking is also accompanied by a notable reduction of the full width at half maximum (FWHM) which reduces from 20 Hz (oscillator 1) and 32 Hz (oscillator 2) to 10 Hz in the frequency locked system. This indicates the improved immunity of the synchronized system to internal phase noise, and agrees with the phase noise reduction shown in several other coupled and self-injection locked oscillators systems [38-41].

Fig. 2a shows the evolution of the synchronization state (coupled vs. uncoupled) of the two oscillators as a function of the injected white noise amplitude for two different values of supply voltage ($V_{DD}$=3V and 4V). It can be observed that in both the cases a minimum threshold noise voltage ($V_{T,NOISE}$) is required to induce frequency locking. Moreover, this critical noise voltage increases with the amplitude of the oscillator output (achieved by increasing $V_{DD}$). This is further elaborated in Fig. 2b which describes the minimum $V_{T,NOISE}$ required as a function of the oscillator signal amplitude. $V_{T,NOISE}$ increases linearly with the oscillator amplitude indicating that a minimum signal to (input) noise ratio (~9:1) is required to induce frequency locking. Further, the relative phase difference (Ø) between the oscillators, frequency locked by noise (Fig. 2c), evolves with the noise amplitude. Here, phase is defined using the relative time difference between the voltage troughs of the waveforms. Each oscillation is considered as a phase change of 2π radians (≡360°) and the relative phase difference is defined as: $\Delta \Phi = (\frac{\Delta t}{T}) \times 360°$ (Δt: minimum time difference between the adjacent troughs of the two oscillators; T: time period). Oscillator 2 (orange in Fig. 1b), with higher stand-alone resonant frequency, initially leads oscillator 1 (green) until they exhibit nearly anti-phase locking; beyond this point, oscillator 2 lags oscillator 1.

**Synchronization of a larger oscillator system.** Further, we also evaluate experimentally the synchronization of a larger system of up to 4 oscillators using white noise injection (Fig. 2d). It can be observed that the noise injection not only enables frequency locking among the oscillators but the critical noise voltage ($V_{T,NOISE}$) reduces with increasing number of oscillators. This can be attributed to the reduced internal phase noise which exhibits an inverse dependence on the number of oscillators in the system[42,43].

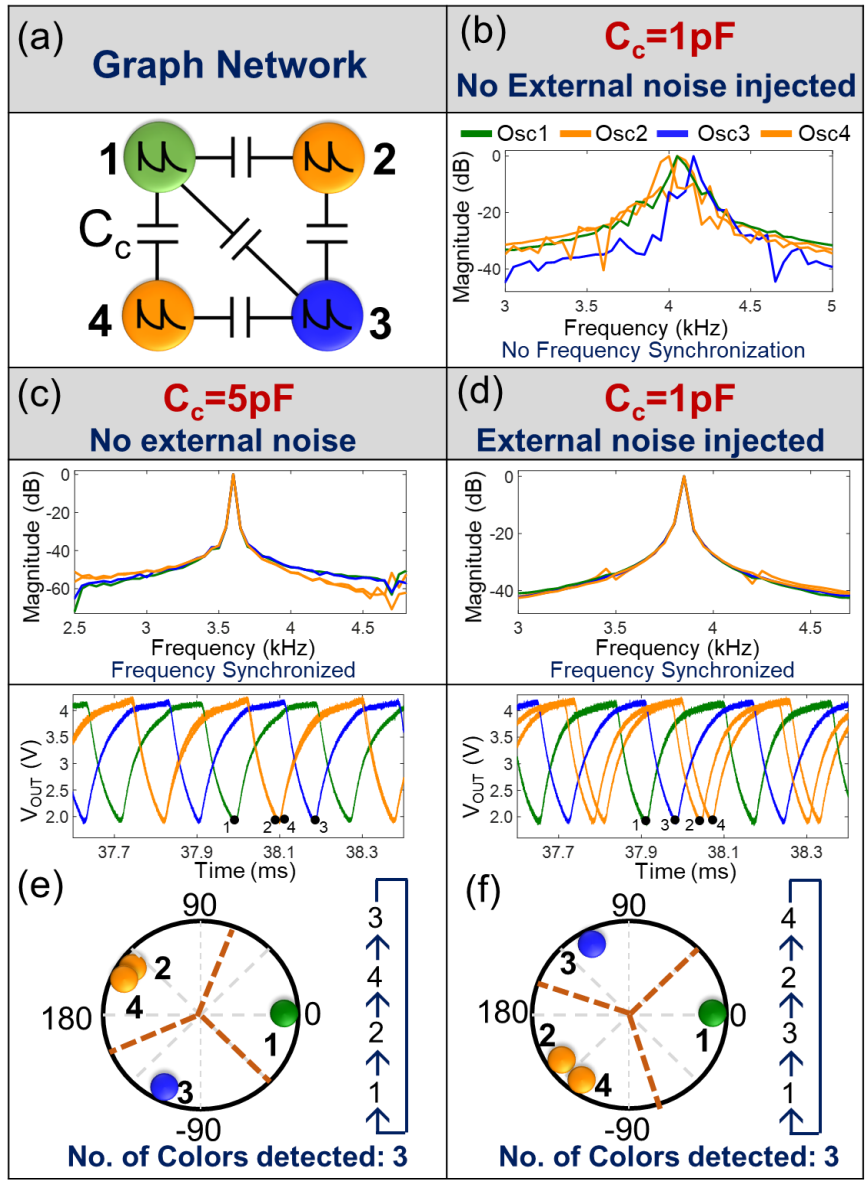

**Figure 3. (a)** Schematic of the oscillator network. Frequency spectrum of the four coupled oscillators with: **(b)** $C_c$=1pF; no frequency locking without injection of noise **(c)** $C_c$ = 5pF, where oscillators exhibit frequency locking owing to the larger value of coupling capacitance; **(d)** $C_c$=1pF, where the noise induced frequency locking is observed. **(e)** and **(f)** Time domain waveforms corresponding to (c) and (d), respectively, along with the phase plots showing the correct phase ordering that gives rise to the optimal graph coloring solution.

**Noise induced synchronization in coupled oscillators.** While the uncoupled oscillator system (synchronized by noise) illustrates how noise promotes synchronization, its effect on the dynamics of a coupled system of oscillators is particularly relevant to computational applications. Consequently, we explore the effect of noise on the frequency locking dynamics of coupled oscillators and their resulting computational properties relevant to solving the graph coloring problem.

The graph coloring problem entails computing the minimum number of colors (labels) required to be assigned to the nodes of a graph such that no two nodes having a common edge are assigned the same color. The problem is NP-hard and is still considered intractable to solve using digital computers, motivating the exploration of alternate approaches. This problem can be elegantly mapped to the oscillator hardware by creating a topologically equivalent coupled oscillator network (graph node ≡ oscillator, and edge ≡ coupling capacitor $C_C$). Subsequently, when the coupling strength is strong enough to frequency lock the oscillators to a common frequency, the resulting phase dynamics exhibit a unique phase ordering such that clusters of nodes without an edge (independent set) and can be assigned the same color, appear consecutively in the circular ordering. The nodes of the same color can then be separated using a simple polynomial time sorting algorithm. Details of this approach have been discussed and demonstrated in our prior work[11]. It is important to emphasize that a critical coupling strength indicated by the magnitude of the coupling capacitance is required to induce frequency locking among the oscillators and observe the desired phase dynamics.

To understand the effect of external noise injection, we first evaluate the minimum coupling strength required to induce synchronization. Figures 3b, c shows the frequency spectrum of the oscillators for the illustrative graph in Fig. 3a for $C_C$= 1pF and 5pF, respectively. In the absence of external noise, a minimum $C_C$= 5pF is required to induce synchronization as shown in Fig. 3c. The corresponding time domain waveforms of the frequency locked oscillators and the phase plots, shown in Fig. 3e, demonstrate a cyclic phase ordering (1, 2, 4, 3, 1...) where independent nodes (i.e. without an edge; 2,4 here) appear consecutively. Using simple post processing, the nodes can be separated into different clusters (= 3, in this problem) of independent sets ({1}, {2,4}, {3}) each of which can be assigned an independent color. Thus, the solution to the graph in Fig. 3a is equal to 3 (colors).

However, the oscillators fail to exhibit frequency locking when $C_C$< 5pF. The minimum coupling strength requirement puts a constraint on the minimum size and area of the coupling element. Since the number of elements in the coupling network of a reconfigurable coupled oscillator based computational platform exhibits a square law dependence (= $P(N,2)$) on the number of oscillators (N), a large footprint for an individual element can dramatically limit the platform's scalability and the reconfigurability.

**Figure 4.** Experimentally observed noise induced synchronization, resulting phase plot and observed graph coloring solutions for various coupled oscillator networks ($C_C$= 1pF).

However, when white noise of appropriate amplitude (180 mV$_{RMS}$) is injected (Fig. 3d), it can be observed that the coupled oscillators not only exhibit frequency locking but also show the optimal phase ordering ({1}, {3}, {2,4}), giving rise to the optimal coloring solution at a significantly lower $C_C$ (= 1pF) (Fig. 3d, f). It is to be noted that even though the relative ordering observed in Fig. 3f is different from Fig. 3e, the ordering is still optimal.

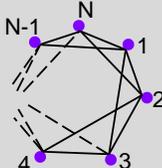

| K=4 Nearest Neighbor Graphs N | Best Solution -$C_C$=50 pF -No noise | Best Solution -$C_C$=0.5 pF -With noise injection | Optimal Solution (=N%3+3) |
|---|---|---|---|
| 8 | 5 | 5 | 5 |
| 16 | 4 | 4 | 4 |
| 32 | 5 | 5 | 5 |
| 64 | 4 | 4 | 4 |

**Figure. 5** Simulations showing coloring of K=4 nearest neighbor graphs of various size (N) using the coupled oscillators, with and without the injection of noise (180 mV$_{RMS}$). It can be observed that the injection of noise reduces the required capacitance to observe frequency locking and the corresponding computation relevant phase properties. $C_C$= 50pF (without noise) and $C_C$= 0.5pF (with noise) were chosen since stable frequency locking at these values for all the analyzed graphs. With $C_C$= 0.5pF, the system does not exhibit synchronization before the injection of noise.

We also experimentally evaluate various coupled oscillator network configurations shown in Fig. 4. In all the configurations, a minimum threshold of $C_C$=5pF is required to induce frequency locking without the external injection of noise. However, the injection of noise helps induce frequency locking at a lower value of $C_C$= 1pF, thus, facilitating a 5× reduction in the minimum required coupling capacitance. Subsequently, this property can be leveraged to proportionally reduce the area required for implementing the coupling capacitor and thus, help the scaling of the area-intensive coupling architecture.

Finally, we evaluate using LT-SPICE simulations, the ability to extend this approach to solving larger graphs. The inset in Fig. 5 shows a schematic of the k=4 nearest neighbor graphs[44] of various sizes up to 64 nodes evaluated here. The same Schmitt trigger oscillator design was considered, and the oscillator dynamics were simulated over a time period of 10ms wherein the system was observed to always attain steady state; multiple runs (>10) were performed for each graph. Subsequently, the relative phase difference among the oscillators is used to construct the phase ordering and compute the coloring solution. Fig. 5 compares the simulated graph coloring solution obtained using the oscillators with larger coupling capacitance ($C_C$=50pF) and without injected noise, with that obtained using smaller coupling capacitance and external noise injection (minimum $V_{NOISE}$ required is between 60 $mV_{RMS}$ (64 oscillators) to 120 $mV_{RMS}$ (8 oscillators)); the optimal solution is also shown for reference. It can be observed that the oscillators with externally injected noise (and lower $C_C$) not only exhibit frequency locking (no frequency locking is observed in the absence of noise when $C_C$=0.5pF) but also produce the same (optimal) solution as those without external noise injection (but higher $C_C$). This further supports the experimental observation that the injection of external noise into the coupled system lowers the minimum coupling capacitance threshold while facilitating the phase dynamics relevant to computation.

**Discussion**

In summary, we elucidate the critical role of injected noise in the synchronization dynamics of uncoupled and coupled (but not frequency locked) oscillators. Moreover, our work demonstrates empirically that noise reduces the minimum coupling strength required to realize the computational properties of the oscillators for solving combinatorial optimization problems such as graph coloring, thus, enabling an additional 'knob' to optimize the implementation and scalability of the area-hungry coupling network in an oscillator platform. Finally, these results also motivate the exploration of the role of noise on the computational performance of other non-Boolean dynamical systems such as spiking neural networks.

**Acknowledgements**

This research was supported in part by the National Science Foundation (Grant No. 1914730).

**Author contributions**

J. V. performed the experiments, simulations and analyzed the data. M. K. B. helped with the performance of the experiments. N. S. supervised the study. J. V., N. S. wrote the manuscript. All authors discussed the results and commented on the manuscript.

**Competing interests**

The authors declare no competing interests.

**Additional information**

Correspondence and requests for materials should be addressed to N.S.

Reprints and permissions information is available at www.nature.com/reprints.